\documentclass[12pt]{article}
\usepackage{eurosym}
\usepackage[left=0.0cm, right=0.0cm, top=0.1in, bottom=0.1in]{geometry}
\usepackage{hyperref}
\usepackage{mathtools}
\usepackage{graphicx}
\usepackage{mathrsfs}
\usepackage{cite}
\newcommand{\bra}[1]{\bigl\langle #1 \bigr|}
\newcommand{\ket}[1]{\bigl| #1 \bigr\rangle}
\begin{document}
	\begin{center}
\textit{\Large Decoherence and quantum steering of accelerated qubit-qutrit system\\}
	\bigskip	
M.Y.Abd-Rabbou$^{a}$ \textit{\footnote{e-mail:m.elmalky@azhar.edu.eg}}, N. Metwally$^{b,c}$\textit{\footnote{e-mail:nmetwally@gmail.com}}, M. M. A. Ahmed $^{a}$,  and A.-S. F. Obada $^{a}$

$^{a}${\footnotesize Mathematics Department, Faculty of Science,
Al-Azhar University, Nasr City 11884, Cairo, Egypt}\\
$^{b}${\footnotesize Math. Dept., College of Science, University of Bahrain, Bahrain.}\\

$^{c}${\footnotesize Department of Mathematics, Aswan University
	Aswan, Sahari 81528, Egypt.}\\
	\end{center}
\begin{abstract}
	The bidirectional steerability between different-size subsystems is discussed for a single parameter accelerated  qubit-qutrit system. The decoherence due to the mixing and acceleration parameters is investigated, where for the total system and the qutrit,  it increases as the mixing parameter increases, while it  decreases  for the qubit. The  non-classical correlations are  quantified by using the local quantum uncertainty, where it increases at large values of the acceleration parameter. The possibility that each  subsystem steers each other is studied, where the behavior of the  steering inequality predicts that the qubit has a large ability to steer the qutrit. The degree of steerability decays gradually when the qubit is accelerated. However, it decays  suddenly when the qutrit or both subsystems are accelerated. The degree of steerability is shown for the qutrit/qubit vanishes at small/large  values of the acceleration. The difference between the degrees of steerability depends on the initial state settings and the size of the accelerated subsystem.

\end{abstract}

\section{Introduction}
\qquad In 1935 Schr{\"o}dinger made an effort to formalize the quantum phenomenon of Einstein-Podolsky- Rosen (EPR) steering, where the possibility that one observer steers or changes the quantum state of another party at a distance by performing appropriately chosen local measurements \cite{1,2}. The strict hierarchy of quantum non-separable states demonstrating that the EPR steering exists in the range between entanglement and Bell non-locality, where the three quantum correlations are inequivalent \cite{3}. So, the Bell non-locality is contained in the steering states, and the steering states are subset in entanglement \cite{4}. In the past few years, quantum steering is implemented in many branches of physics in experimental and information-theoretic tasks, such as self-testing of quantum states \cite{5}, secure teleportation \cite{6}, randomness generation \cite{7}, and quantum key distribution \cite{8}.  A nondegenerate optical parametric oscillator has accomplished the violation of continuous variable EPR steering inequality \cite{9}. The phenomenon of EPR steering is detected in the optical system experimentally by using entangled two-photon states  \cite{10}. Theoretically, quantum steering and its steerability can be reported for different quantum system via violating the steering inequalities. For example, it has been discussed for bipartite two-qubit X-state \cite{11,12,13}, optical cavity according to multipartite system\cite{14}, Heisenberg chain model \cite{15,16}, two-level or three-level detectors \cite{17,18}. As quantum systems interact with several surrounding environments, it was essential to indicate the effect of these surroundings on the steering correlation \cite{19,20}. For instance, the efficacy of relativistic motion, the noisy channel, finite temperature, non-Markovian environment, and a cavity optomechanical system on the steering have been discussed \cite{12,21,22,23,24}. Mathematically, the degree of steerability and steering inequality have been diagnosed via different relations such as, Heisenberg uncertainty principle \cite{25}, steering witnesses\cite{26}, the standard geometric Bell inequalities \cite{27,28}, and the maximal violation of the Clauser-Horne-Shimony-Holt inequality \cite{29,30}.

The optimal steering inequality for a pair of arbitrary discrete observables is obtained by Walborn et al. \cite{25} as,
\begin{equation}\label{1}
	H(R^B|R^A)+	H(Q^B|Q^A)\geq \log_2(\Omega^B)\qquad \text{with}, \Omega^B\equiv \max_{i,j}\{|\langle R^B|Q^B\rangle|^2\},
\end{equation}
  where $H(R^B|R^A)=H(\rho^{AB})- H(\rho^{A})$ is the conditional entropy, and $\{|R^B\rangle \}$ and $\{|Q^B\rangle \}$ are eigenstates of the observables $ R^B  $ and $ Q^B $. By employing the Pauli spin matrices, one can find the optimal steering inequality as following  \cite{31},
  \begin{equation}
  	H(S_x)+H(S_y)+H(S_z)\geq \gamma_s,
  \end{equation}
where $\gamma_s $= 2, 3 for the qubit and the qutrit respectively.

In the other hand, investigating the effect of the Unruh framework is one of the significance studies in relativistic quantum processing, where the quantum regimes are basically non-inertial \cite{32}. The  Unruh framework effects on the quantum steering for the maximal entangled mixed state  has been studied \cite{33}. The dynamical behaviour of quantum steering between two-modes of  Dirac fields interact locally with thermal baths in the non-inertial frame has been investigated \cite{34}. However, influence of the acceleration  on the quantum correlation \cite{35,36}, quantum coherence \cite{37}, estimation degree \cite{38,39}, and the quantumness via Wigner distribution \cite{40} have been discussed.

Our motivation in this study is to formalize a general form of entropic uncertainty steering inequality for the one-parameter family of qubit state (2D) interacting locally with qutrit state (3D), whether the qubit (small dimension) steers the qutrit (large dimension) or qutrit steers qubit. Also, the Unruh framework effects are taken into account,  where the qubit and the qutrit are accelerated individually, or simultaneously.  Meanwhile, the degree of steerability of the accelerated system, and the relation between the decoherence and steerability are investigated.

This article is organized as follows; In Sec.(\ref{ss2}), we describe the  one family parameter qubit-qutrit system and the acceleration process, where it is assumed that,  either the qubit, qutrit, or both of them are accelerated. The decoherence due to the mixing parameter and the acceleration process is discussed in Sec.(\ref{Decoherence}).  The amounts of non-classical correlations are quantified by using the local quantum uncertainty in Sec.(\ref{LQU}). Sec.(\ref{DS}) is devoted to investigate the   bidirectional steerability process between the accelerated subsystems. Finally, we summarize our results in Sec.(\ref{Con}).

 \section{The Model.}\label{ss2}

  Let us consider that  a system  of one parameter type, that consists of a qubit system interacting locally with a qutrit system.  In the computational basis, the system may be written as\cite{41},
 \begin{equation}\label{2.1}
 	\begin{split}
 		\rho^{qt}(p)=\frac{p}{2}&\left\lbrace |00\rangle\langle 00|+ |01\rangle\langle 01|+|11\rangle\langle 11|+|12\rangle\langle 12| +|12\rangle\langle 00|+|00\rangle\langle 12| \right\rbrace \\&+\frac{1-2p}{2}\left\lbrace |02\rangle\langle 02|+ |02\rangle\langle 10|+|10\rangle\langle 02|+|10\rangle\langle 10| \right\rbrace,
 	\end{split}
 \end{equation}
 where $ 0\leq p \leq 0.5$.
  It is assumed that,  the qubit subsystem is initially accelerated uniformly  whilst the qutrit subsystem  is in a fixed frame.  The computational basis $|0_{q}\rangle $ and $|1_{q}\rangle $ of the qubit state in the Minkowski coordinates are transformed into the Rindler coordinates as \cite{42}.
 \begin{equation}\label{2.4}
 	\begin{split}
 		|0_q\rangle=\cos r_q |0_q\rangle_{I} |0_q\rangle_{II} + \sin r_q |1_q\rangle_{I} |1_q\rangle_{II} , \quad |1_q\rangle= |1_q\rangle_{I} |0_k\rangle_{II},
 	\end{split}
 \end{equation}
where  $r\in[0,\pi/4]$ is the acceleration parameter. Meanwhile, the computational basis set  $ \{\ket{0},\ket{1},~\ket{2}\} $ of the qutrit state are transformed from the Minkowski coordinates into the Rindler  coordinates as \cite{43},
 \begin{equation}\label{2.5}
 	\begin{split}
 		&|0_t\rangle=\cos^2 r_t |0\rangle_{I} |0\rangle_{II}+e^{i \phi} \cos r_t \sin r_t \left(|1\rangle_{I} |2\rangle_{II} +|2\rangle_{I} |1\rangle_{II}\right) + e^{2i \phi} \sin^2 r_t |\uparrow \downarrow\rangle_{I} |\uparrow\downarrow\rangle_{II}, \\&
 		|1_t\rangle= \cos r_t |1\rangle_{I} |0 \rangle_{II}+ e^{i \phi} \sin r_t |\uparrow\downarrow\rangle_{I} |1\rangle_{II},\\&
 		|2_t\rangle= \cos r_t |2\rangle_{I}|0\rangle_{II}-e^{i \phi} \sin r_t |\uparrow\downarrow\rangle_{I} |2\rangle_{II},
 	\end{split}
 \end{equation}
 where $ |\uparrow\downarrow\rangle $ is pair state.
 Subsequently, we consider that either the qubit and the qutrit are accelerated individually or the bipartite system is accelerated simultaneously.  The output accelerated systems of the three cases in region $I$ may be written as,
 \begin{equation}\label{d1}
 	\begin{split}
 		\rho ^{q,t,qt}_{acc}=&\rho^{q,t,qt}_{11}|00\rangle\langle 00|+ \rho^{q,t,qt}_{22}|01\rangle\langle 01|+\rho^{q,t,qt}_{33}|02\rangle\langle 02|+\rho^{q,t,qt}_{44}|10\rangle\langle 10|+\rho^{q,t,qt}_{55}|11\rangle\langle 11|\\&+\rho^{q,t,qt}_{66}|12\rangle\langle 12|+\rho^{t,qt}_{77}|0 \uparrow\downarrow\rangle\langle 0 \uparrow\downarrow|+\rho^{t,qt}_{88}|1 \uparrow\downarrow \rangle\langle 1 \uparrow\downarrow|+ \big[\rho^{q,t,qt}_{34}|02\rangle\langle 10|\\& +\rho^{q,t,qt}_{16}|00\rangle\langle 12|+\rho^{t,qt}_{57}|11 \rangle\langle 0 \uparrow\downarrow|+\rho^{t,qt}_{28}|01 \rangle\langle 1 \uparrow\downarrow|+h.c.\big],
 	\end{split}
 \end{equation}
 where the superscripts $q,t,qt$ indicates to  the three cases of accelerated qubit, qutrit and qubit-qutrit system, respectively.  The  non-zero elements when only the qubit system is accelerated in Eq.(\ref{d1}) are given by,
 \begin{equation}\label{2.6}
 	\begin{split}
 		&\rho^{q}_{11}=\frac{p}{2}c^2= \rho^{q}_{22} ,\quad \rho^{q}_{33}=\frac{1-2 p}{2}c^2 ,\quad \rho^{q}_{44}=\frac{p}{2}s^2+ \frac{1-2p}{2}, \quad \rho^{q}_{55}=\frac{p}{2}(s^2+ 1) ,\\&
 		\rho^{q}_{66}=\frac{1-2p}{2}s^2+ \frac{p}{2},\quad \rho^{q}_{16}=\frac{p}{2}c = \rho^{q}_{61} ,\quad \rho^{q}_{34}=\frac{1-2p}{2}c = \rho^{q}_{43} ,\quad
 	\end{split}
 \end{equation}
 where $ c= \cos r $ , $ s=\sin r $.
Likewise, the non-zero elements of the output accelerated system $ \rho^{t}_{acc} $ when only  the qutrit is separately accelerated are obtained by,
 \begin{equation}\label{2.7}
 	\begin{split}
 		&\rho^{t}_{11}=\frac{p}{2}c^4 , \quad \rho^{t}_{22}=c^2\frac{p}{2}(s^2+ 1) , \quad \rho^{t}_{33}=\frac{c^2}{2}(p s^2 -2p+ 1) ,  \quad  \rho^{t}_{44}=\frac{1}{2}(1-2p) c^4, \\&
 		\rho^{t}_{55}=\frac{c^2}{2} ((1-2p)s^2+ p )=\rho^{t}_{66}, \quad \rho^{t}_{77}=\frac{s^2}{2}(p s^2-p+1) ,\quad \rho^{t}_{88}=s^2 (\frac{1-2p}{2}s^2+p) ,  \\&  \rho^{t}_{16}= \frac{p}{2} c^3  ,\ \rho^{t}_{28}=-\frac{p}{2}c\ s^2  ,  \quad \rho^{t}_{34}=\frac{1-2p}{2}c^3,\quad \rho^{t}_{57}=\frac{2p-1}{2}c \ s^2.
 	\end{split}
 \end{equation}

 Finally, the non-zero elements when the total system is  accelerated  are given by,
 \begin{equation}\label{2.8}
 	\begin{split}
 		&\rho^{qt}_{11}=c^2 \rho^{t}_{11} , \quad \rho^{qt}_{22}=c^2 \rho^{t}_{22}  , \quad \rho^{qt}_{33}=c^2\rho^{t}_{33}  , \quad  \rho^{qt}_{44}=\rho^{t}_{55} +s^2 \rho^{t}_{11}   , \quad \rho^{qt}_{55}= \rho^{t}_{55} +s^2 \rho^{t}_{22},\\&
 		\rho^{qt}_{66}= \rho^{t}_{66}+s^2 \rho^{t}_{33}, \quad \rho^{qt}_{77}=c^2\rho^{t}_{77}  ,\quad\rho^{qt}_{88}= \rho^{t}_{88} +s^2 \rho^{t}_{77}, \quad \rho^{qt}_{16}=c  \rho^{t}_{16},\quad
 		\rho^{qt}_{28}=c  \rho^{t}_{28},\\& \qquad  \rho^{qt}_{57}=c  \rho^{t}_{57},\qquad \rho^{qt}_{34}=c  \rho^{t}_{34},
 	\end{split}
 \end{equation}
 Our motivation is to quantify the amount of the steerability from the qubit to the qutrit and vice versa.

\section{Degree of decoherence:}\label{Decoherence}
The degree of quantum decoherence has been constructed in terms of linear entropy and von Neumann entropy.  \cite{44,45}. Indeed, the quantum decoherence phenomenon means the off-diagonal elements which generate the coherence between quantum regimes in the density operator are terminated. The decoherence degree is dissipated automatically for the mixed  and the pure states when they interact with their environment, where the pure states have a minimum value of decoherence (zero value). The degree of quantum decoherence is based on linear entropy which is defined by,
\begin{equation}
	\mathcal{D}^{qt,q,t}=1-Tr[\rho^{qt,q,t}]^2,
\end{equation}
where $ qt,q, \text{and}\ t $ refer to the density operator of the qubit-qutrit, qubit, and qutrit states, respectively.

\begin{figure}[h!]
	\centering
	\includegraphics[width=0.4\linewidth, height=3.5cm]{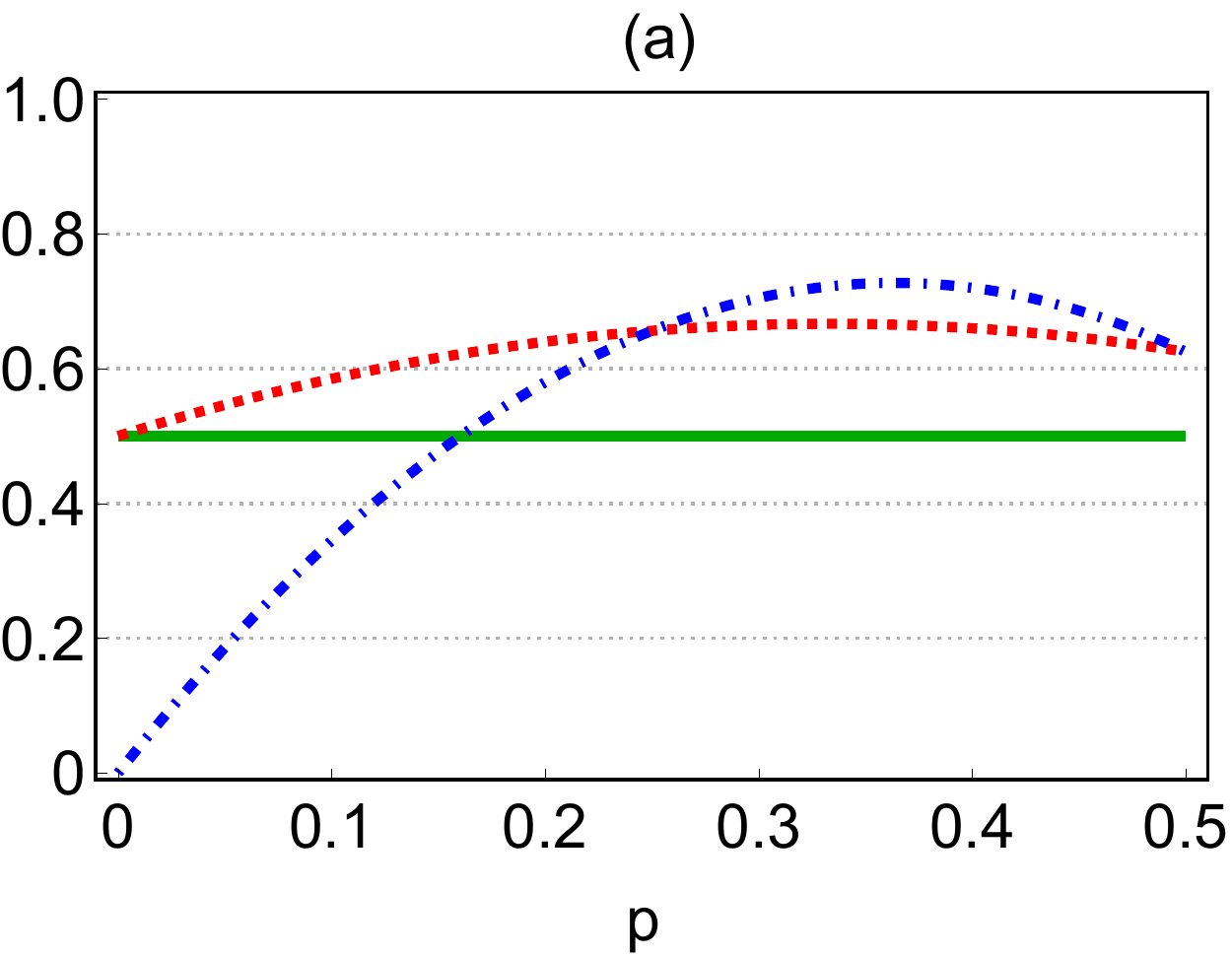}
	\includegraphics[width=0.4\linewidth, height=3.5cm]{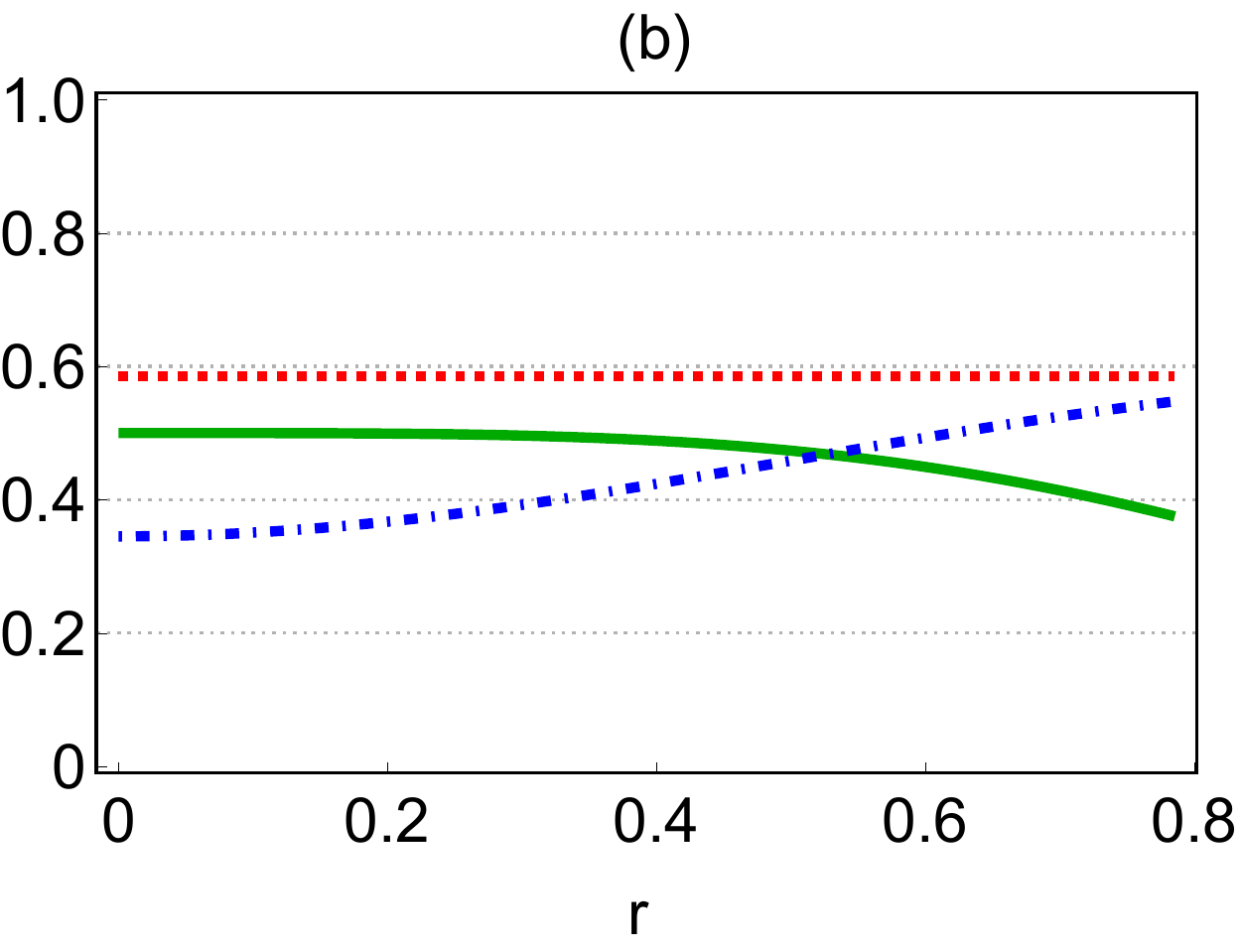}
	\includegraphics[width=0.4\linewidth, height=3.5cm]{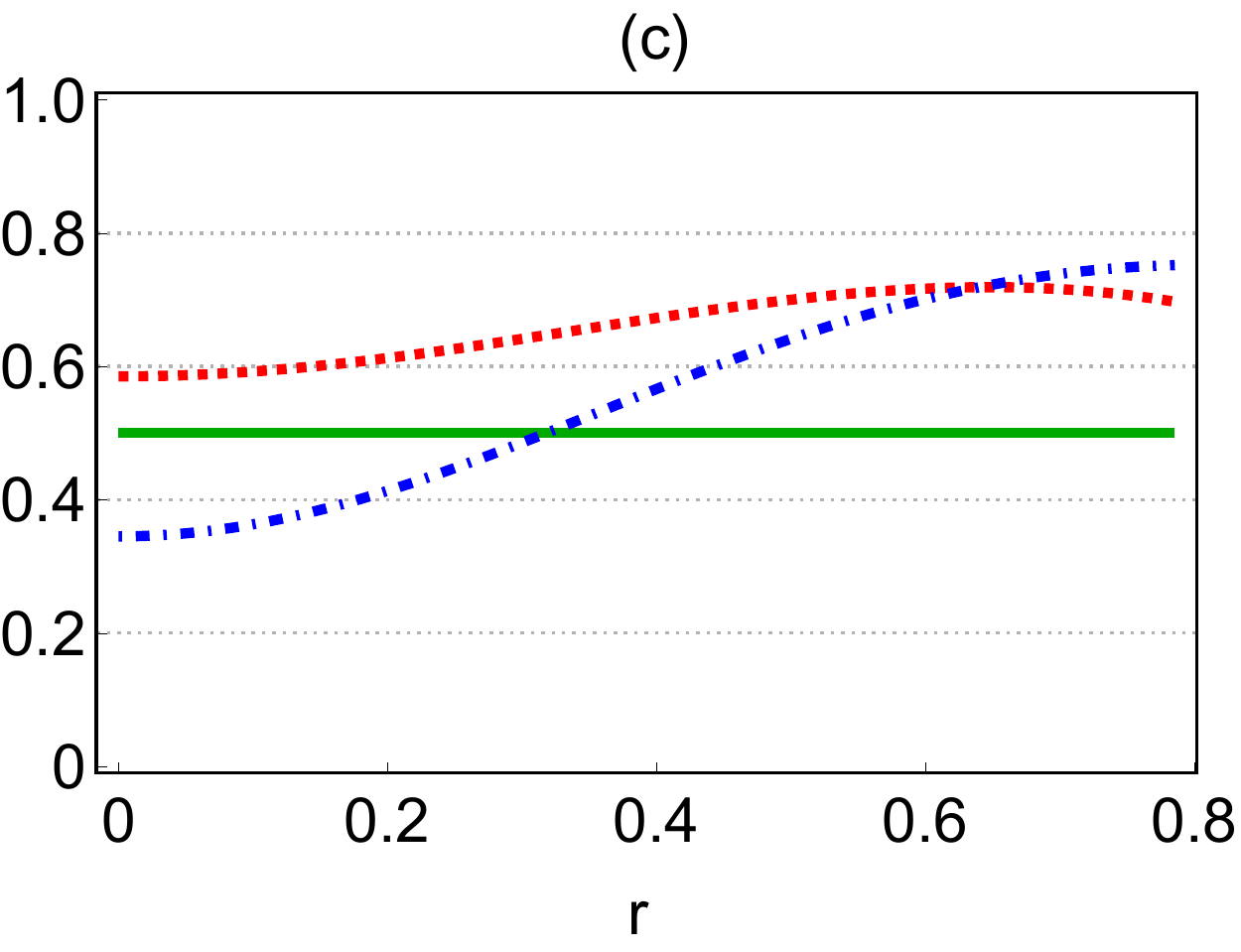}
	\includegraphics[width=0.4\linewidth, height=3.5cm]{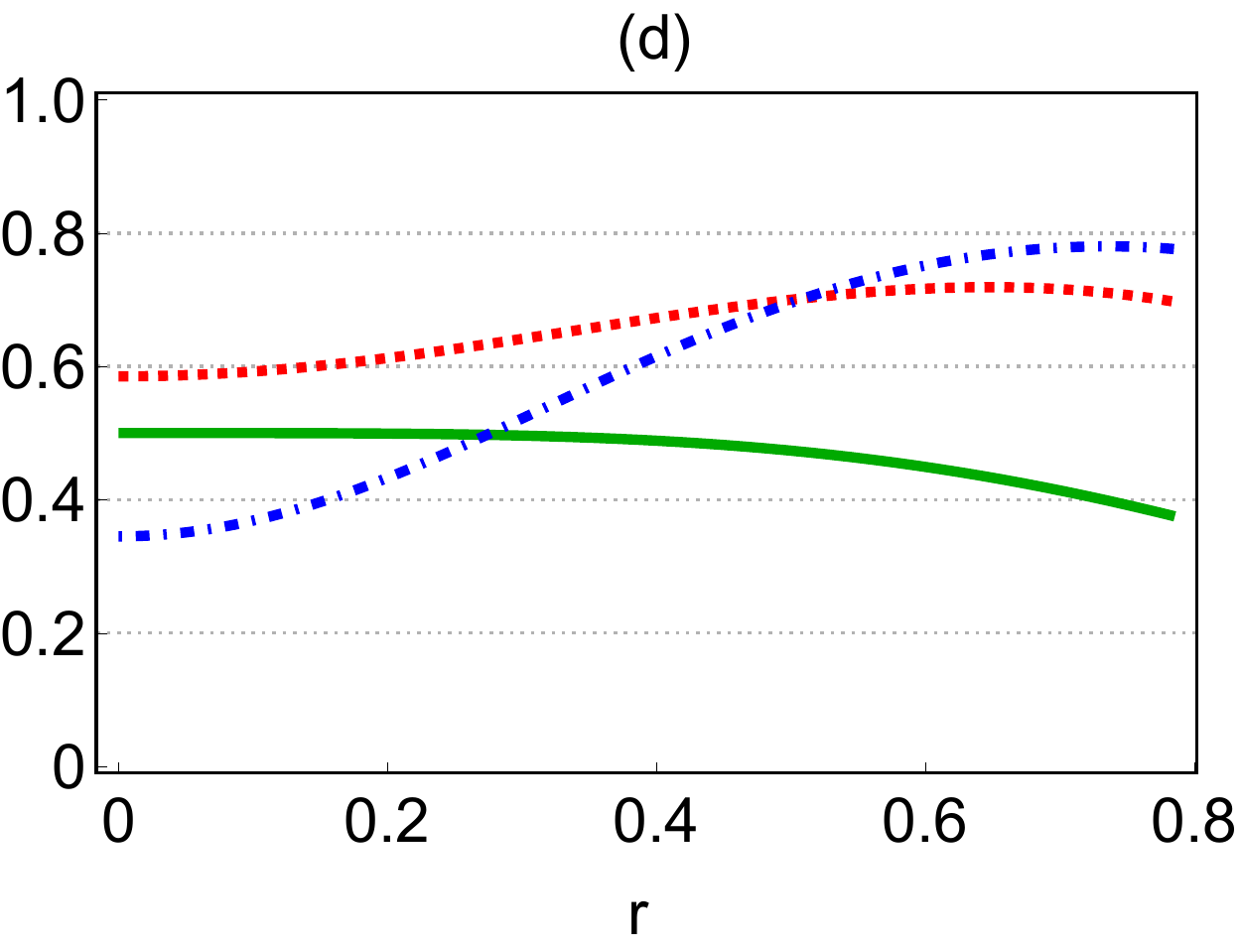}
	\caption{Quantum decoherence  $ \rho_{AB} $ (blue dot-dash-curve), decoherence $ \rho_{A} $ (green solid-curve),  decoherence $ \rho_{B} $(red dash-curve) (a) zero acceleration, (b) accelerated qubit with p=0.1,  (c)accelerated qutrit with p=0.1, and (d) both qubit and qutrit are accelerated with p=0.1.}
	\label{f5}
\end{figure}

It is worth studying the amount of decoherence that arises from the mixture parameter $p$ and the acceleration $r$. Fig.(\ref{f5}) displays the  behavior of the decoherence for the non-accelerated/accelerated qubit-qutrit system. For the non- accelerated system, the behavior of the decoherence is depicted in Fig.(\ref{f5}a). It is clear that, at $p=0$, the initial qubit-qutrit system $\rho^{qt}$ is maximum entangled state and the system is a decoherence free. However, as $p$ increases, the initial state $\rho^{qt}$ loses its coherence  due to the mixture of a non pure  state. The decoherence increases gradually as the mixture parameter $p$ increases. Also, the qubit state of this initial state $\rho^q=tr_t\{\rho^{qt}\}=\frac{1}{2}I_{2\times 2}$, namely is independent of the  mixture parameter $p$. Therefore, as it is displayed from  Fig.(\ref{f5}a),  there is no decoherence depicted on the initial qubit system. The initial qutrit system namely $\rho^t=tr_q\{\rho^{qt}\}=p\ket{1}\bra{1}+\frac{1-p}{2}(\ket{0}\bra{0}+\ket{2}\bra{2})$,   depends on the mixture parameter $p$. However, as $p$ increases, the decoherence that displayed for the initial qutrit increases gradually to reach its maximum values at $p=0.5$.

Fig.(\ref{f5}b) displays the behavior of the decoherence on the total initial system and its composite systems, where it is assumed that only the qubit is accelerated and  we set  the mixing parameter $p=0.1$. As it is displayed from this figure, at $r=0$, the predicted decoherence is due to the mixing parameter. However,  the   decoherence of the total system $\rho^{qt}$ increases gradually as $r$ increases, where the maximum decoherence is depicted at large acceleration, i.e., $r\approx0.8$. Similarly the decoherence of the qutrit system   due only the  mixing parameter, where its value does not change as the qubit is accelerated. The effect of the acceleration on the degree of decoherence is displayed on the accelerated qubit, where for $r>0.4$, the decoherence decreases.

The behavior of the decoherence for  the three states when only the qutrit is accelerated is shown in Fig.(\ref{f5}c). It is clear that the decoherence on the qutrit system that due to the acceleration is depicted at $r>0.2$. However, as  $r$ increases, the qubit system loses its coherence gradually. As it is expected the accelerating process on the qutrit has no effect on the coherence degree of the qubit, where the displayed decoherence only due to the mixing parameter. The coherent degree on the qubit-qutrit system is displayed clearly, where the decoherence of $\rho^{qt}$ is depicted at small accelerations and increases fast as $r$ increases. Moreover, the upper bounds of this decoherence of the total state is smaller than that displayed for the qutrit for any $r<0.6$. At further values of $r>0.6$, the decoherence  depicted for $\rho^{qt}$ is larger than that displayed for $\rho^{t}$.

Fig.(\ref{f5}d) describes the effect of the accelerated  qubit-qutrit on the decoherence of the initial total system and its composite states. It is clear that, at $r=0$, the decoherence on the three states duo to the mixing parameter, where we set $p=0.1$. However, the initial decoherence of the qutrit is the largest one, while it is the smallest one for the total state $\rho^{qt}$. As one increases the acceleration of the qubit and the qutrit, the decoherence of the total state and the qutrit increases gradually, while the degree of coherence of the qubit improves.

From Fig.(\ref{f5}), one may conclude that, for  this suggested qubit-qutrit  system, the accelerating process on the qubit improves its  coherence. The decoherence due to the mixing parameter depicted for the qutrit is larger than that displayed for the total state. If the qutrit is accelerated, the total state and the reduced state of the qutrit lose their coherence as the acceleration increases.

\section{Local quantum uncertainty}\label{LQU}

In this section, we investigate the local/non local behavior of the accelerated initial qubit-qutrit state. For this aim, we consider the local quantum uncertainty, LQU as a predictor of the non-locality and quantifier of the amount of the non-classical correlations NCCs.
If the operator $ O $ is an observable we need to quantify the  minimum amount of LQU in state $ \rho_{A,B} $. Thus, the LQU with respect to the qutrit $ A $ is defined as a minimum Wigner–Yanase skew information, which given by,
\begin{equation}
	U_A(\rho_{A,B})=\min\limits_{O_A}[\mathcal{I}(\rho_{A,B},O_A)],
\end{equation}
where $ \mathcal{I}(\rho_{A,B},O_A):= \frac{-1}{2} Tr[\rho_{A,B},O_A]^2$, and $ O_A:= O_A\otimes I_b $. For $ 2\otimes 3 $ the expression of LQU is written as,
\begin{equation}
	U_A(\rho_{A,B})=1-\max[\Gamma_1,\Gamma_2,\Gamma_3]
\end{equation}
where $ \Gamma_i $ are the eigenvalues of the $ 3 \times 3 $ symmetric matrix $ \Xi_{AB} $ with entries $ (\Xi_{AB})_{ij}=Tr[\sqrt{\rho_{A,B}}(S_i\otimes I_3)\sqrt{\rho_{A,B}}(S_j\otimes I_3)] $, and $ S_i,S_j $ are the Pauli operators for the qubit system.
\begin{figure}[!t]
\centering
	\includegraphics[width=0.32\linewidth, height=4cm]{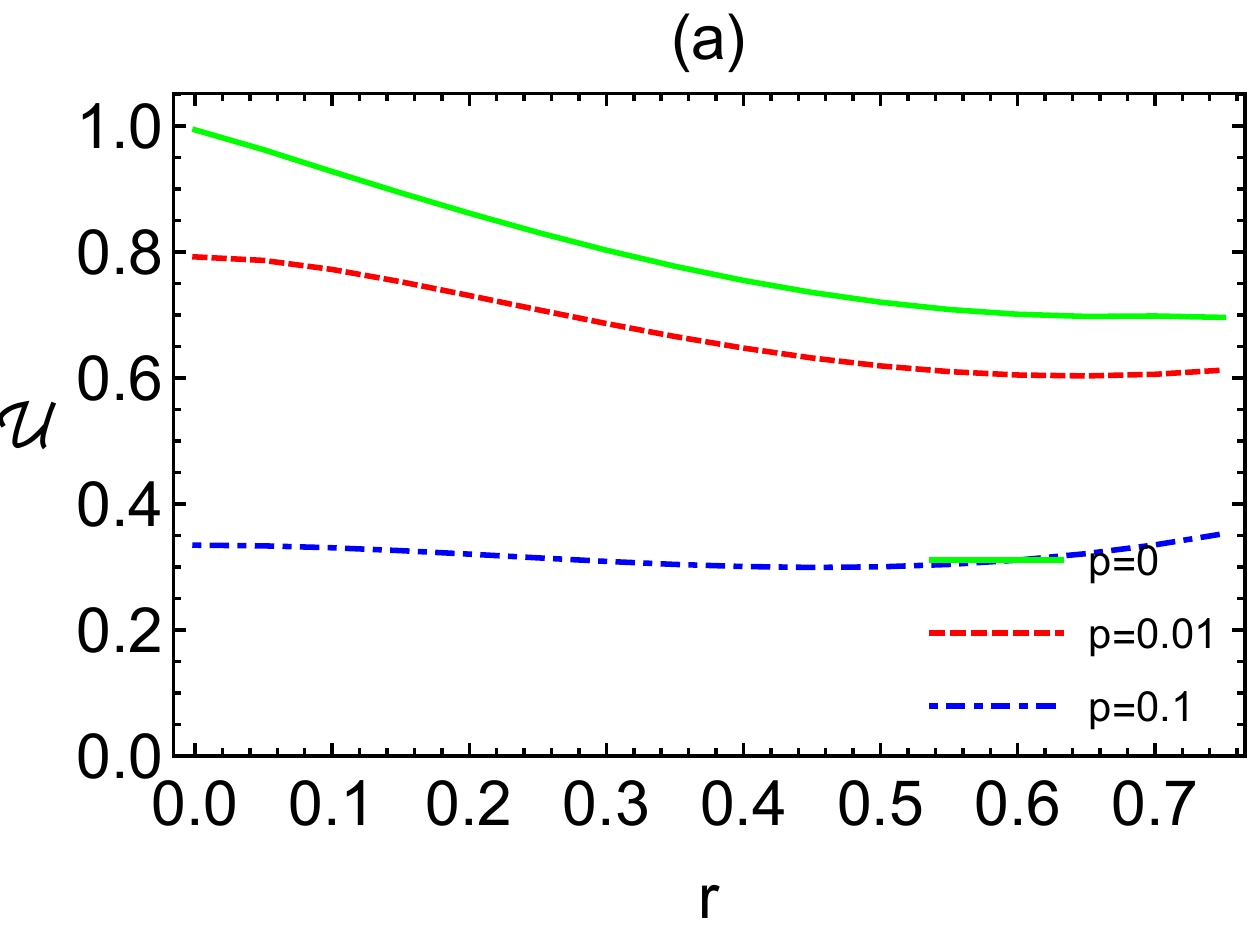}~\quad
	\includegraphics[width=0.32\linewidth, height=4cm]{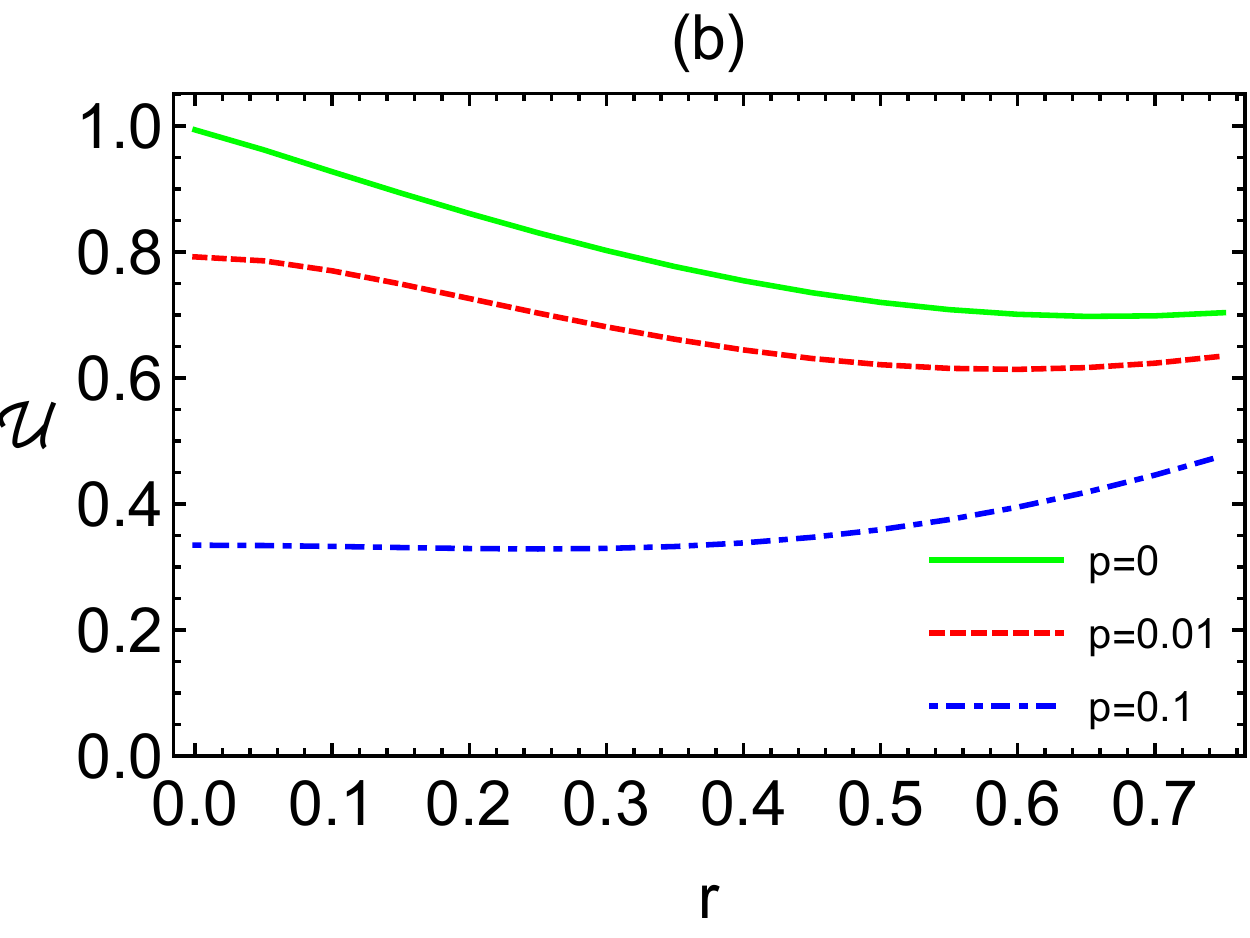}~\quad
	\includegraphics[width=0.32\linewidth, height=4cm]{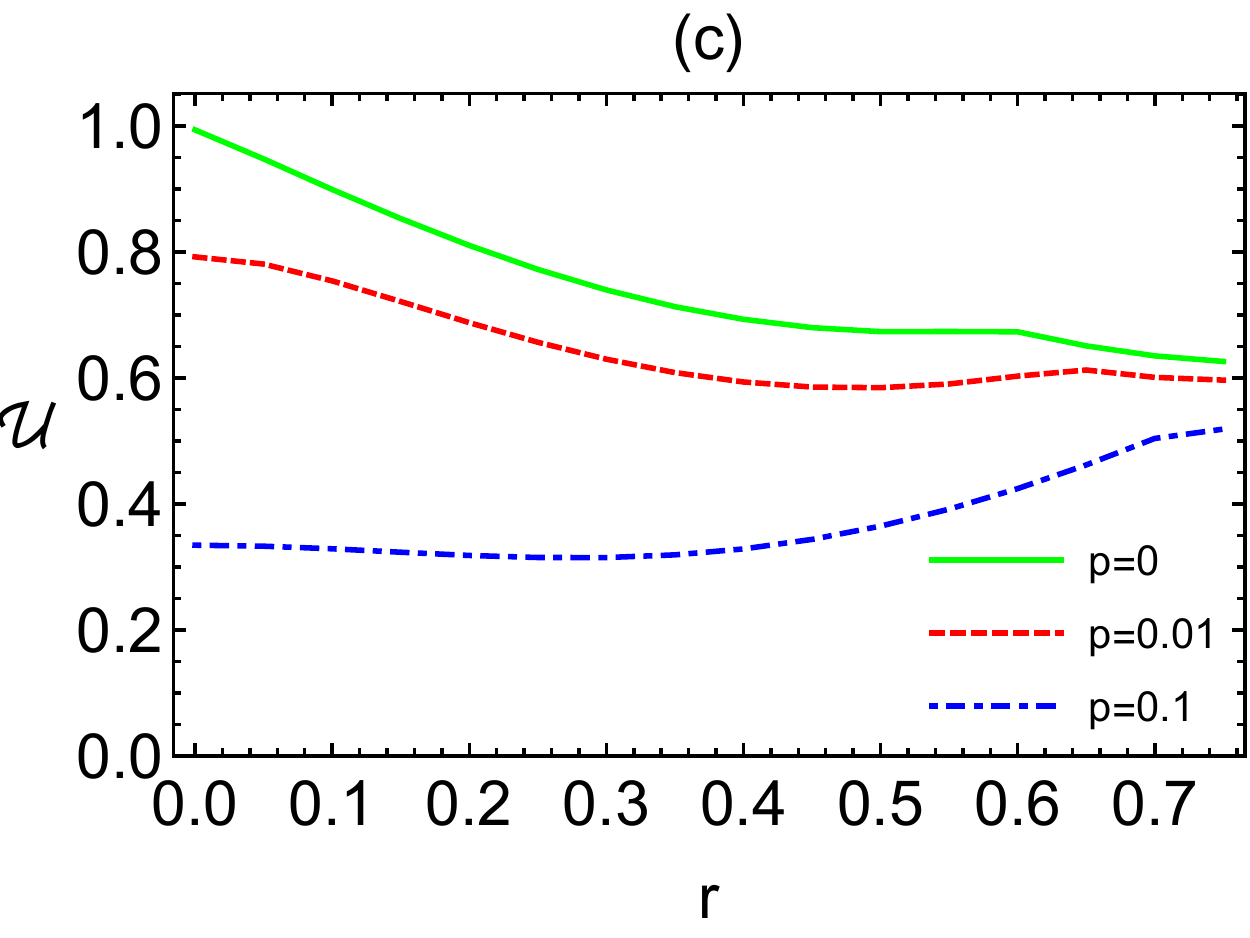}
	\caption{ Local quantum uncertainty $\mathcal{u}$  (a) accelerated qubit,  (b)accelerated qutrit, and (c) both qubit and qutrit are accelerated.}
	\label{QLU}
\end{figure}

 Fig.(\ref{QLU}), shows the behavior of LQU ($\mathcal{u}$), where it is assumed that either, the qubit, qutrit or both of them are accelerated. Fig.(\ref{QLU}a), displays the behavior of $\mathcal{u}$ when only the qubit is accelerated and different values of the mixing parameter. It is clear that at $p=0$, namely the initial state is maximally entangled state, the QLU  decreases gradually as the acceleration increases. However,  as one increases the  value of the mixing parameter, the decreasing rate of the LQU increases. At further values of $p$, the decoherence due to the acceleration is very small, where LQU is almost invariant and the amount of  quantum correlations is slightly affected as $r$ increases. In Fig.(\ref{QLU}b), we investigate the effect of the accelerated qutrit on the behavior of the local quantum uncertainty. The behavior is similar to that displayed in Fig.(\ref{QLU}a), but the deceasing rate of QLU  is slightly smaller than that displayed in  Fig.(\ref{QLU}a) and it is faintly  decreases as the acceleration increases. Moreover as it is displayed from Fig.(\ref{QLU}c), the local quantum uncertainty increases as the acceleration increases at large values of the mixing parameter.

 \section{Degree of Steerability:}\label{DS}
In this section, we investigate the steerability from the local qubit to the local qutrit in the absence or presence of the acceleration. In terms of the steering inequality (\ref{1}), the amount of steering from the local qubit system (A) to the local qutrit system (B) can be identified via violating of the following inequality,
\begin{equation}\label{sab}
	S_{AB}=H(S_x^B|S_x^A)+H(S_y^B|S_y^A)+H(S_z^B|S_z^A)\geq 3.
\end{equation}
Hence, the steerability by the qubit's measurements to render $ S_{AB}\in[0,1] $, is expressed by,
\begin{equation}
	S^{A\longrightarrow B}=\max\bigg\{0,\frac{S_{AB}-3}{S_{max}-3}\bigg\},
\end{equation}
where the $ S_{max}=4 $. In this case, and the denominator is a normalization factor. Also the inequality of steering from B to A is given by,
\begin{equation}\label{sba}
	S_{BA}=H(S_x^A|S_x^B)+H(S_y^A|S_y^B)+H(S_z^A|S_z^B)\geq 2.
\end{equation}
Likewise, the steerability by the qutrit's measurements to make $ S_{BA}\in[0,1] $ is defined by,
\begin{equation}
	S^{B\longrightarrow A}=\max\bigg\{0,\frac{S_{BA}-2}{S_{max}-2}\bigg\},
\end{equation}
with $ S_{max}=3 $. In the computational basis $\{\ket{0},~\ket{1}\}$, and $\{\ket{0},~\ket{1},~\ket{2}\}$, the operators $\hat{S}_i^{A,B}, i=x,y,z$ that describe the local qubit and qutrit for the users A and B, respectively, are defined by \cite{PhysRevA.93.062126},
\begin{equation}
	\begin{split}
		&\hat{S}_x^A=|0\rangle \langle 1|+|1\rangle \langle 0|, \ \hat{S}_y^A=-i\big(|0\rangle \langle 1|-|1\rangle \langle 0|\big), \ \hat{S}_z^A=|0\rangle \langle 0|-|1\rangle \langle 1|,\\ &
		\hat{S}_x^B=-i \big(|1\rangle \langle 2|-|2\rangle \langle 1||\big), \ \hat{S}_y^B=i \big(|0\rangle \langle 2|-|2\rangle \langle 0|\big),\ \hat{S}_z^B=-i \big(|0\rangle \langle 1|-|1\rangle \langle 0|\big),
	\end{split}
\end{equation}
According to Eqs. (\ref{sab}) and (\ref{d1}),  the inequality of the steering from the qubit to the qutrit is giving by,
\begin{equation}
	\begin{split}
			\mathcal{I}_{AB}&=(1-a)\log_2(1-a)+a \log_2 a+b \log_2 b+\frac{c^+}{2}\log_2(c^+)+ \frac{c^-}{2}\log_2(c^-) + \\&
			+(\rho^{q,t,qt}_{11}+\rho^{q,t,qt}_{22}) \log_2  2^5 (\rho^{q,t,qt}_{11}+\rho^{q,t,qt}_{22})+ \rho^{q,t,qt}_{33} \log_2  2^5 \rho^{q,t,qt}_{33} +\rho^{q,t,qt}_{66} \log_2 2^5 \rho^{q,t,qt}_{66}\\&+(\rho^{q,t,qt}_{44}+\rho^{q,t,qt}_{55}) \log_2  2^5 (\rho^{q,t,qt}_{44}+\rho^{q,t,qt}_{55}) -\frac{d^-}{2} \log_2 (d^-) -\frac{d^+}{2} \log_2 (d^+) \leq3
	\end{split}
\end{equation}
where,
\begin{equation*}
	\begin{split}
		&a= \rho^{q,t,qt}_{11}+\rho^{q,t,qt}_{44}, \quad b= \rho^{q,t,qt}_{22}+\rho^{q,t,qt}_{55}, \quad
		c^\pm=1- b\pm 2(\rho^{q,t,qt}_{16}+\rho^{q,t,qt}_{34}),  \\& \quad \text{and} \qquad d^\pm=1\pm \big( \rho^{q,t,qt}_{11}+\rho^{q,t,qt}_{22}+\rho^{q,t,qt}_{33}-\rho^{q,t,qt}_{44}-\rho^{q,t,qt}_{55}-\rho^{q,t,qt}_{66}\big).
	\end{split}
\end{equation*}
On the other hand, the steering inequality from the qutrit to qubit is given by,
\begin{equation}
	\begin{split}
		\mathcal{I}_{BA}&=\frac{c^+}{2}\log_2(c^+)+ \frac{c^-}{2}\log_2(c^-)
		+(\rho^{q,t,qt}_{11}+\rho^{q,t,qt}_{22}) \log_2  4(\rho^{q,t,qt}_{11}+ \rho^{q,t,qt}_{22}) + \rho^{q,t,qt}_{33} \log_2  4\rho^{q,t,qt}_{33} \\&+\rho^{q,t,qt}_{66} \log_2  4\rho^{q,t,qt}_{66}+(\rho^{q,t,qt}_{44}+\rho^{q,t,qt}_{55}) \log_2  4(\rho^{q,t,qt}_{44}+\rho^{q,t,qt}_{55}) - (1-b) \log_2 (1-b) \\&-(1-g) \log_2 (1-g) -g \log_2 g \leq2,
	\end{split}
\end{equation}
with, $ g= \rho^{q,t,qt}_{33}+\rho^{q,t,qt}_{66} $. Hereinafter, we introduce a comparative study for the steerability between the two subsystems.

\begin{figure}[!h]
	\centering
	\includegraphics[width=0.45\linewidth, height=3.5cm]{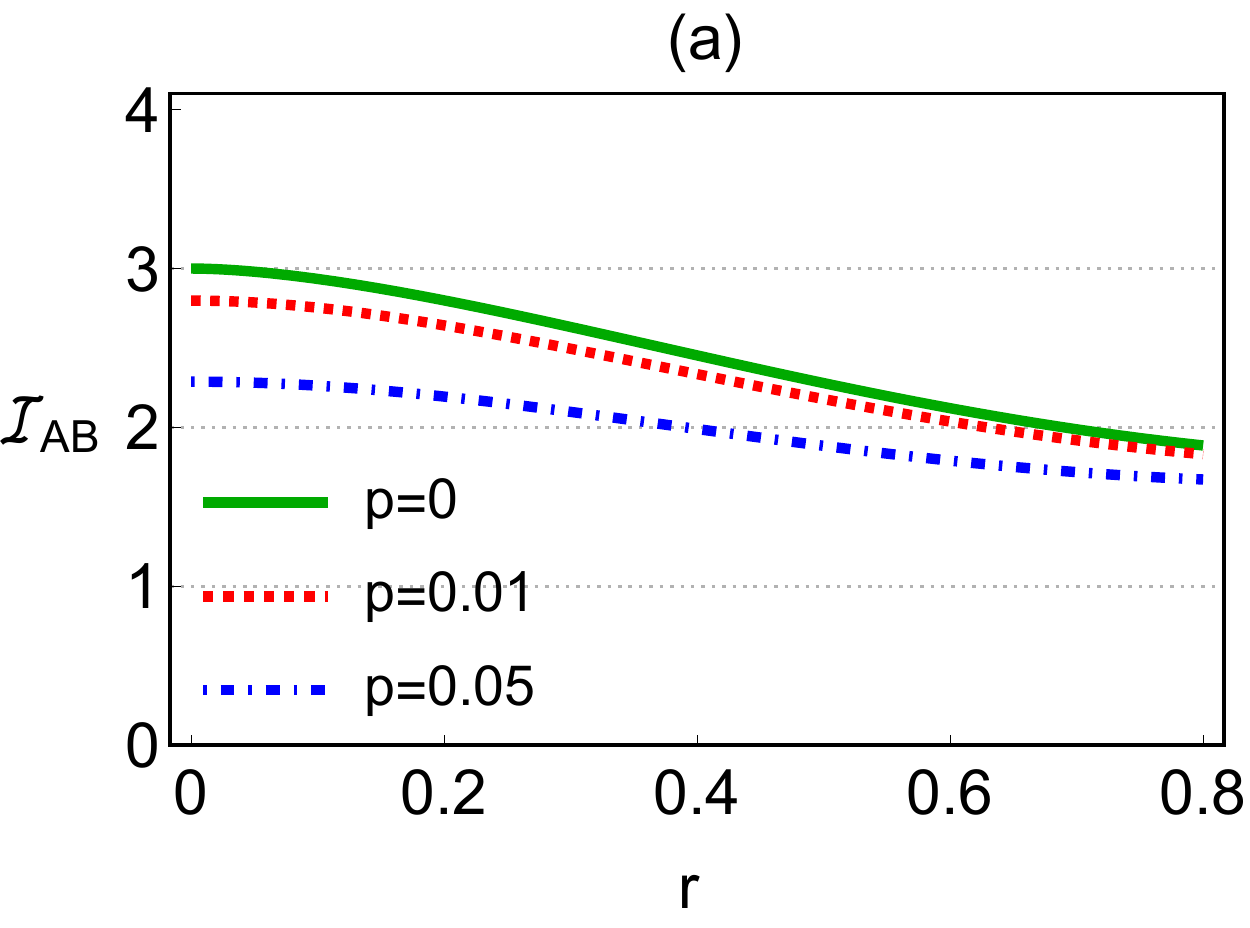}
	\includegraphics[width=0.45\linewidth, height=3.5cm]{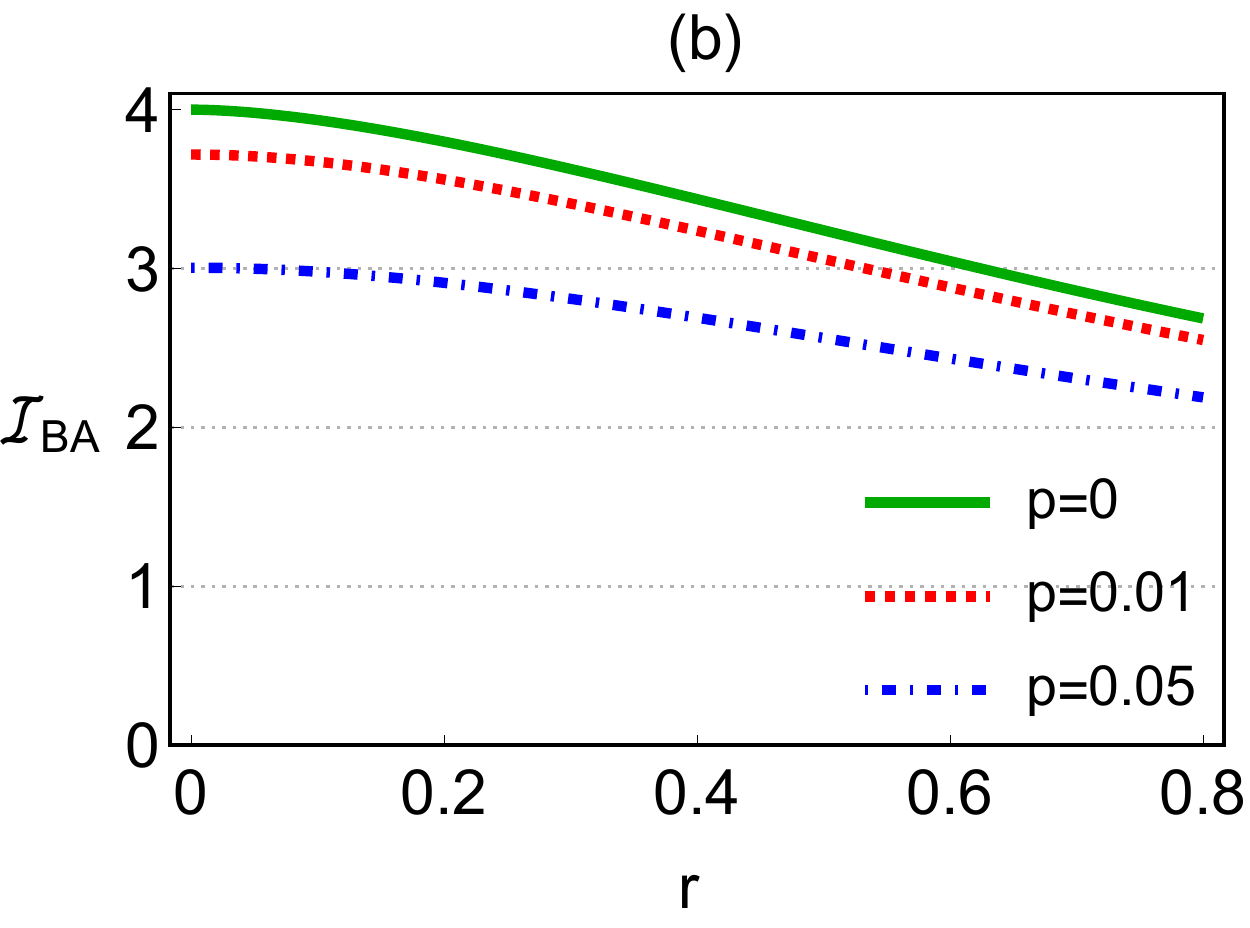}
	\includegraphics[width=0.45\linewidth, height=3.5cm]{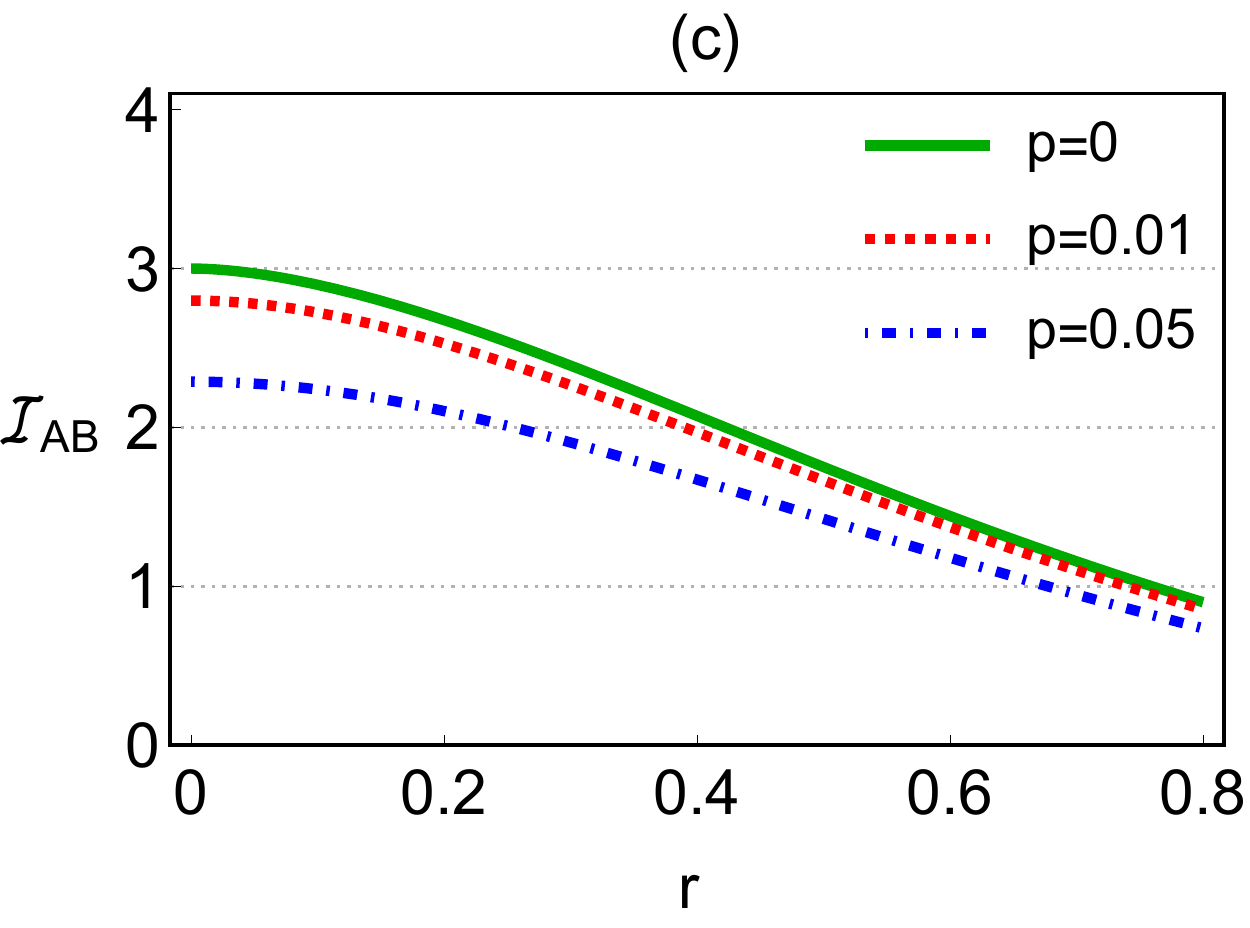}
	\includegraphics[width=0.45\linewidth, height=3.5cm]{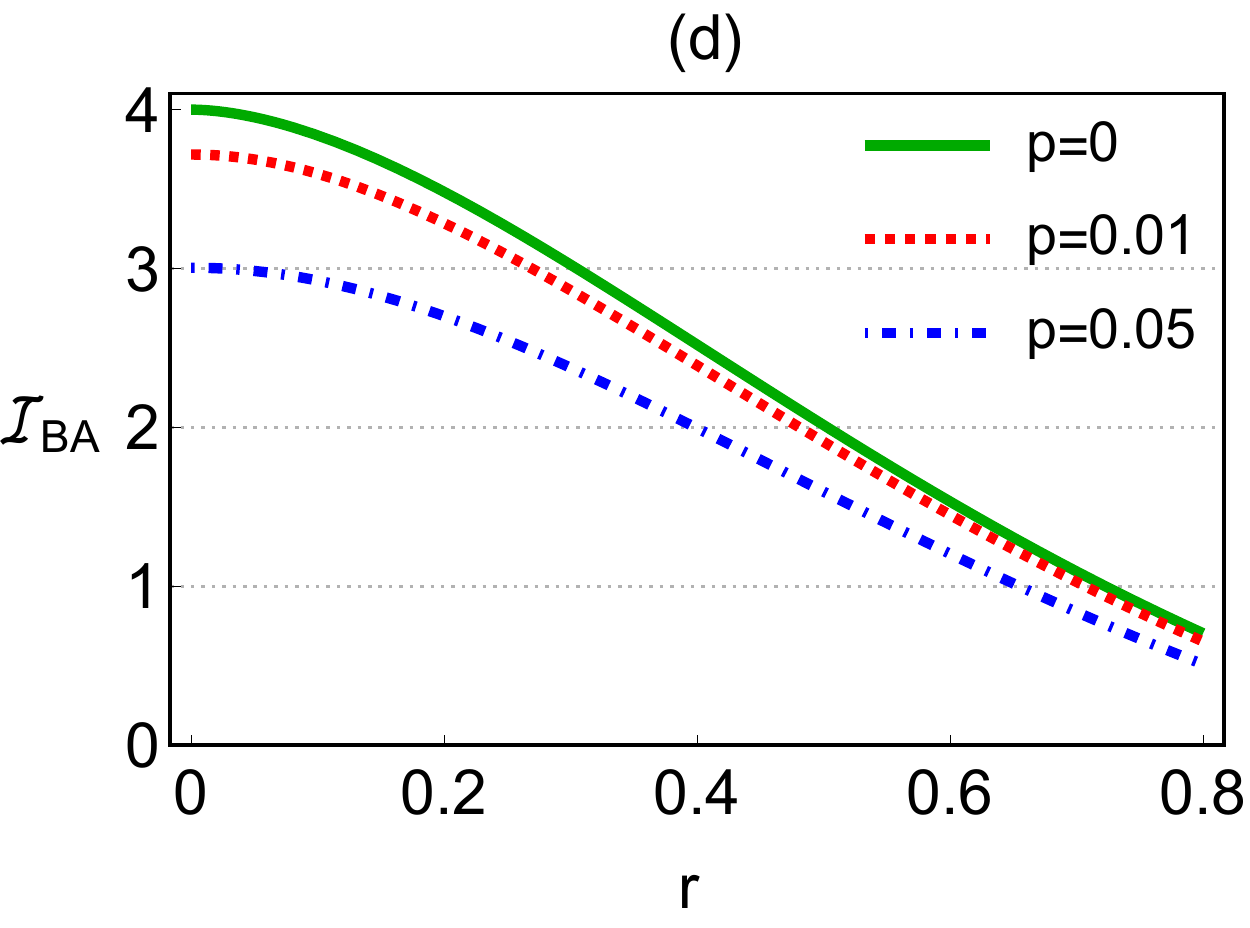}
	\includegraphics[width=0.45\linewidth, height=3.5cm]{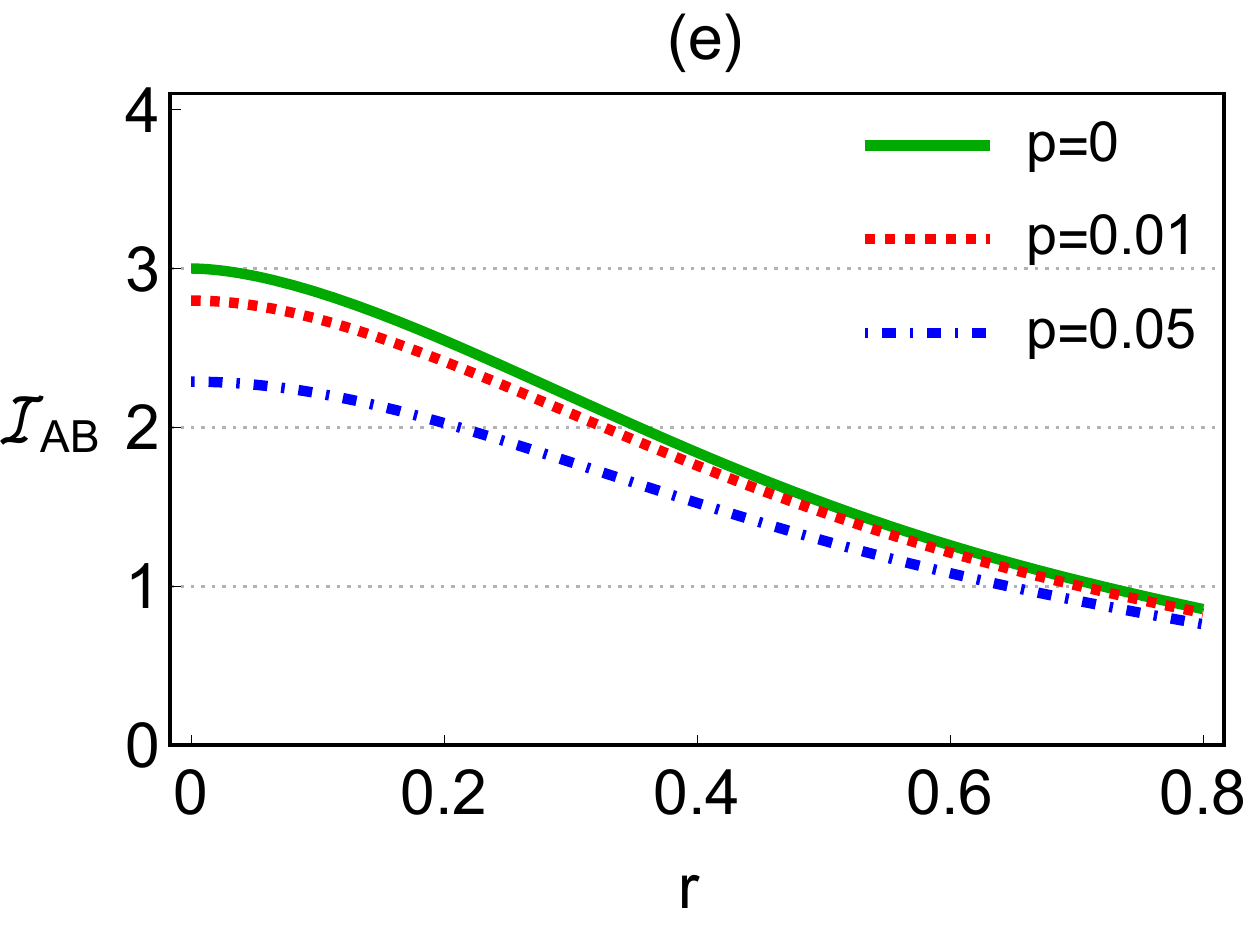}
	\includegraphics[width=0.45\linewidth, height=3.5cm]{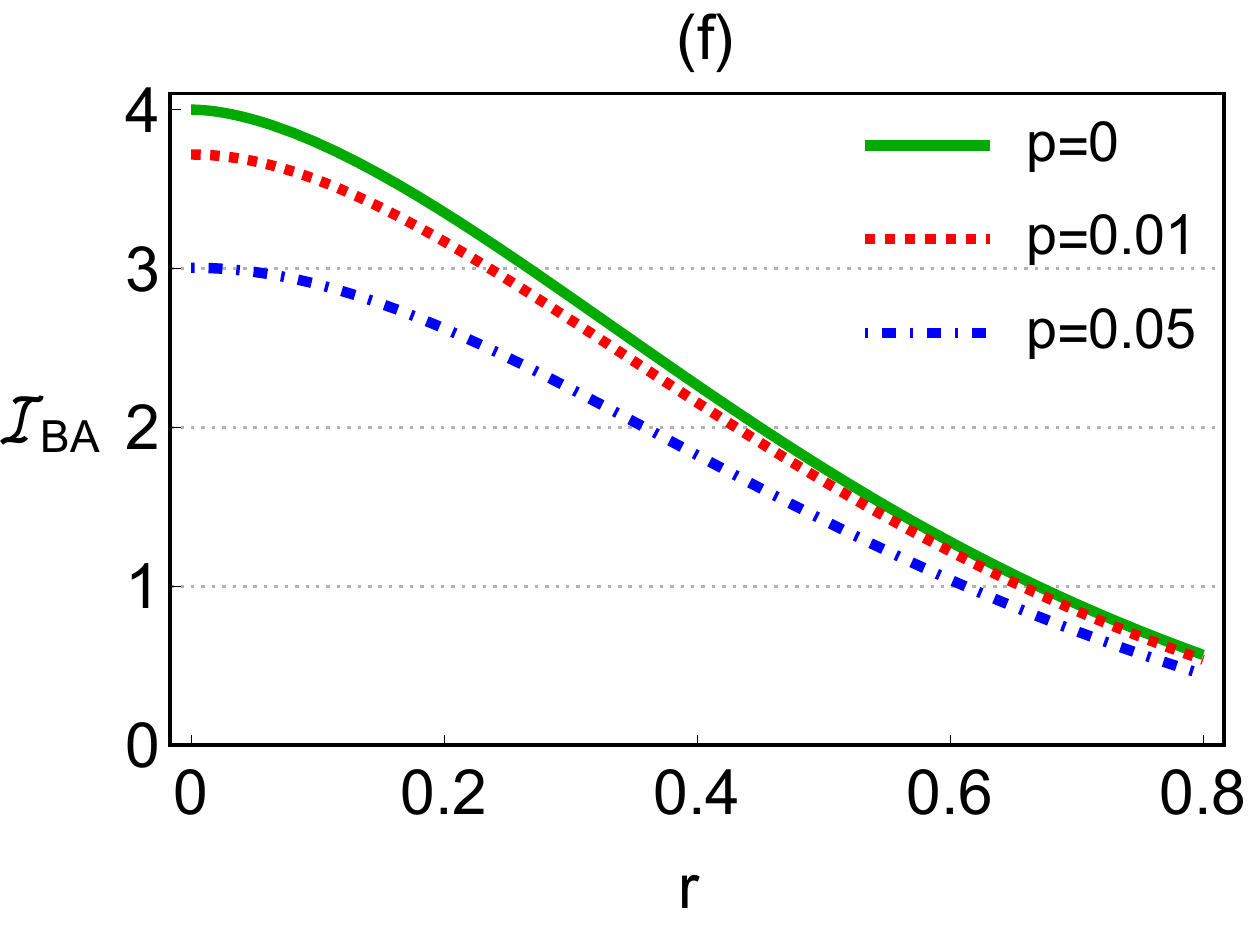}
	\caption{ steering inequality if the qubit is accelerated   (a) steering from qubit to the qutrit $\mathcal{I}_{AB}$,  (b) steering from the qutrit to the qubit $\mathcal{I}_{BA}$. (c,d) the same as (a,b) respectively, but the qubit is accelerated. (e,f)  the same as (a,b) respectively, but the qubit and the qutrit are accelerated.}
	\label{figS}
\end{figure}

 The possibility that either the qubit or the qubit steers each other is  displayed  in Fig(\ref{figS}), for different initial states.
 In general, the steering inequality decreases gradually as the acceleration parameter increases. The decreasing rate depends on the initial settings of the accelerated state. The behavior of the steering inequalities  shows the smallest decay depicted for the maximum entangled state, i.e., the mixing parameter $p=0$.  As one increases the mixing parameter, namely increasing the degree of decoherence, the steering inequality decreases fast as the acceleration increases.  As it is displayed from Figs. (\ref{figS}a,b) and (\ref{figS}c,d), the size of the accelerated particle  (qubit/qutrit) has a remarkable effect on the behavior of the steering inequalities. It is  shown that, if the qutrit is accelerated, the depicted  decreasing rate   is larger than that displayed if the qubit is accelerated. Moreover, these inequalities decrease faster when both  subsystems (qubit and the qutrit) are accelerated.
 On the other hand the steering of qutrit via the qubit and vise versa depends on which object is accelerated.

\begin{figure}[h!]
	\centering
	\includegraphics[width=0.315\linewidth, height=3.5cm]{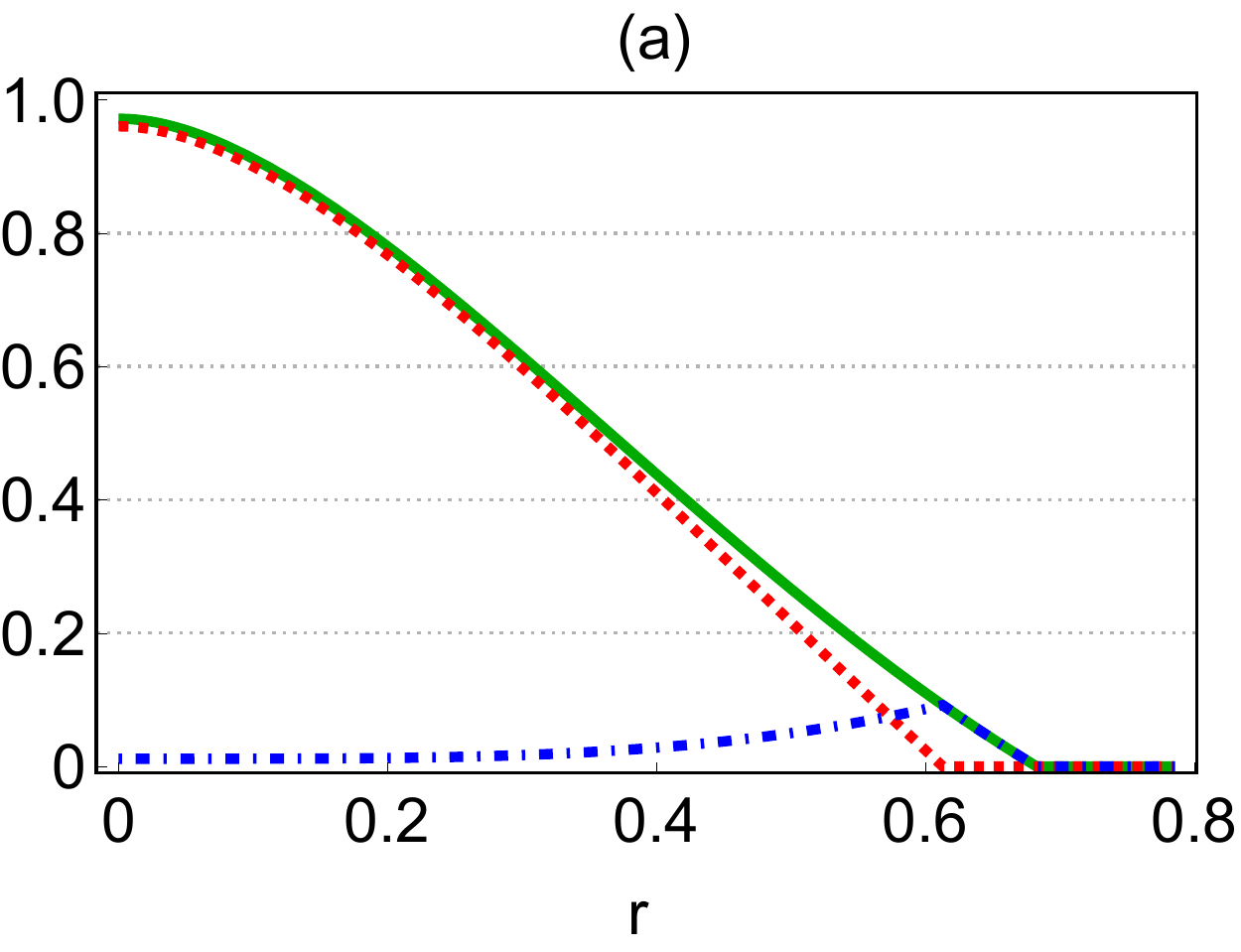}
	\includegraphics[width=0.315\linewidth, height=3.5cm]{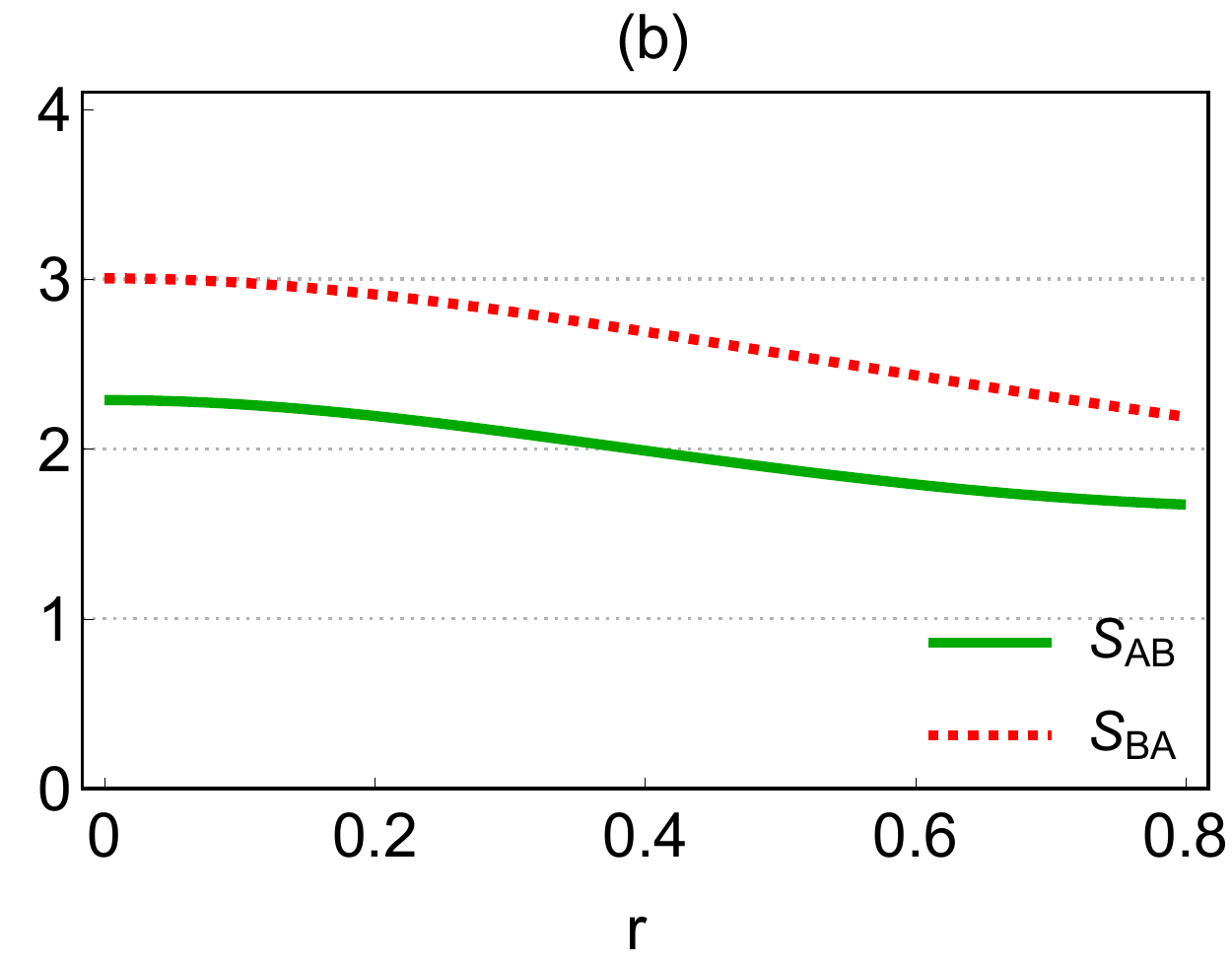}
	\includegraphics[width=0.315\linewidth, height=3.5cm]{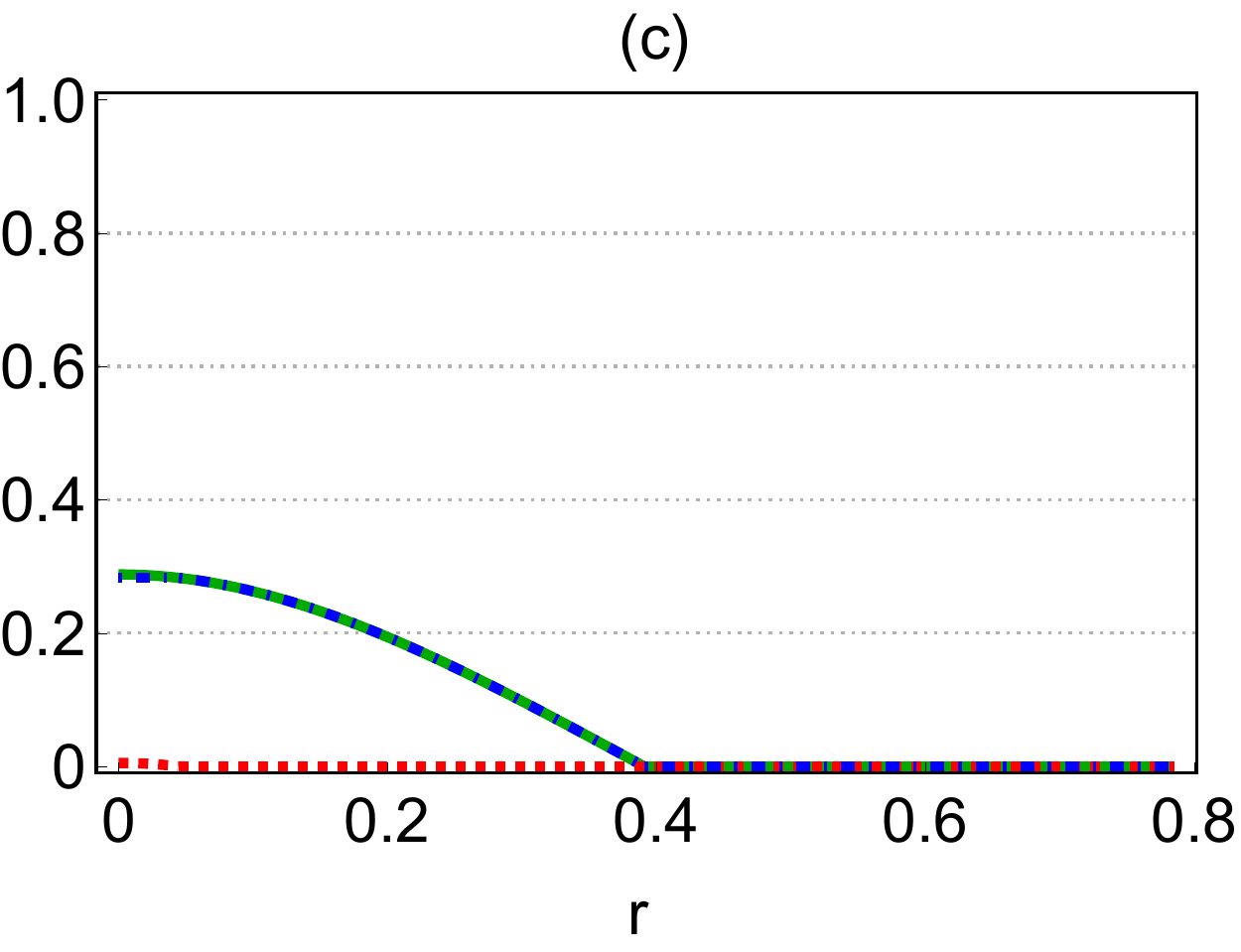}
	\caption{The steerability if the qubit is accelerated, $\mathcal{S}_{AB}$ (green curve), $\mathcal{S}_{BA}$ (red curve), and $|\mathcal{S}_{AB}-\mathcal{S}_{BA} |$ (blue curve) (a) $p=0$, (b)$p=0.01$, and (c)$p=0.05$}
	\label{figDS-q}
\end{figure}

The degree of steerability that the qubit steers the qutrit  and vise versa is displayed in Fig.(\ref{figDS-q}), where different values of the mixing parameter are considered, and only the qubit is accelerated.  This  figure shows that, the degree of steerability decreases as the  degree of  decoherence increases, namely either $p$ or $r$ increases.
 The possibility that, the qubit steers the qutrit, $\mathcal{S}_{AB}$ is larger than that displayed for $\mathcal{S}_{BA}$.  The steerability that, the qubit steers the qutrit vanishes at large values of $ r $ compared with that for $\mathcal{S}_{BA}$. The difference between the degree of steerabilities $|
\mathcal{S}_{AB}-\mathcal{S}_{BA}|$  increases as the mixing and the accelerating parameters increase.

\begin{figure}[h!]
	\centering
	\includegraphics[width=0.315\linewidth, height=3.5cm]{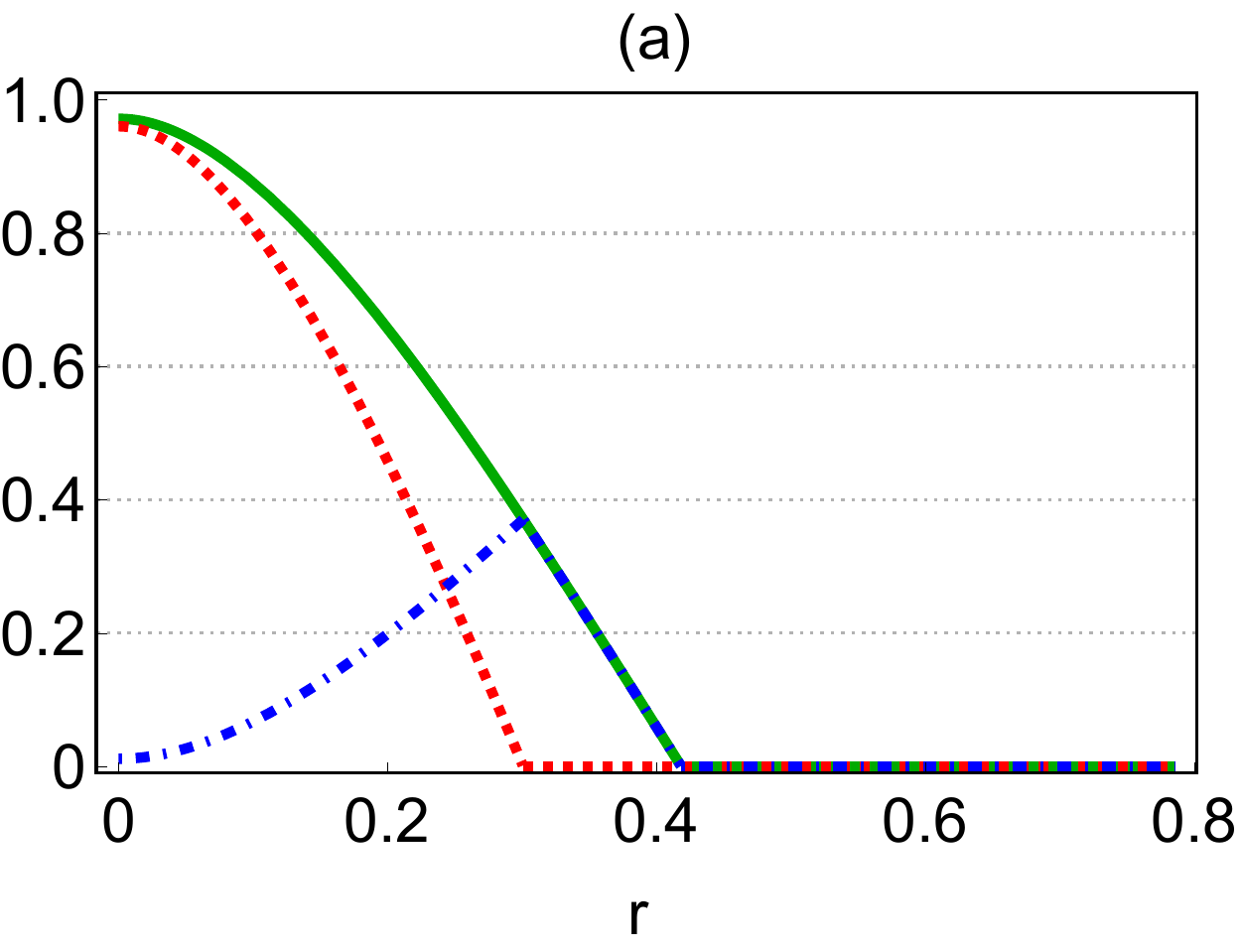}
	\includegraphics[width=0.315\linewidth, height=3.5cm]{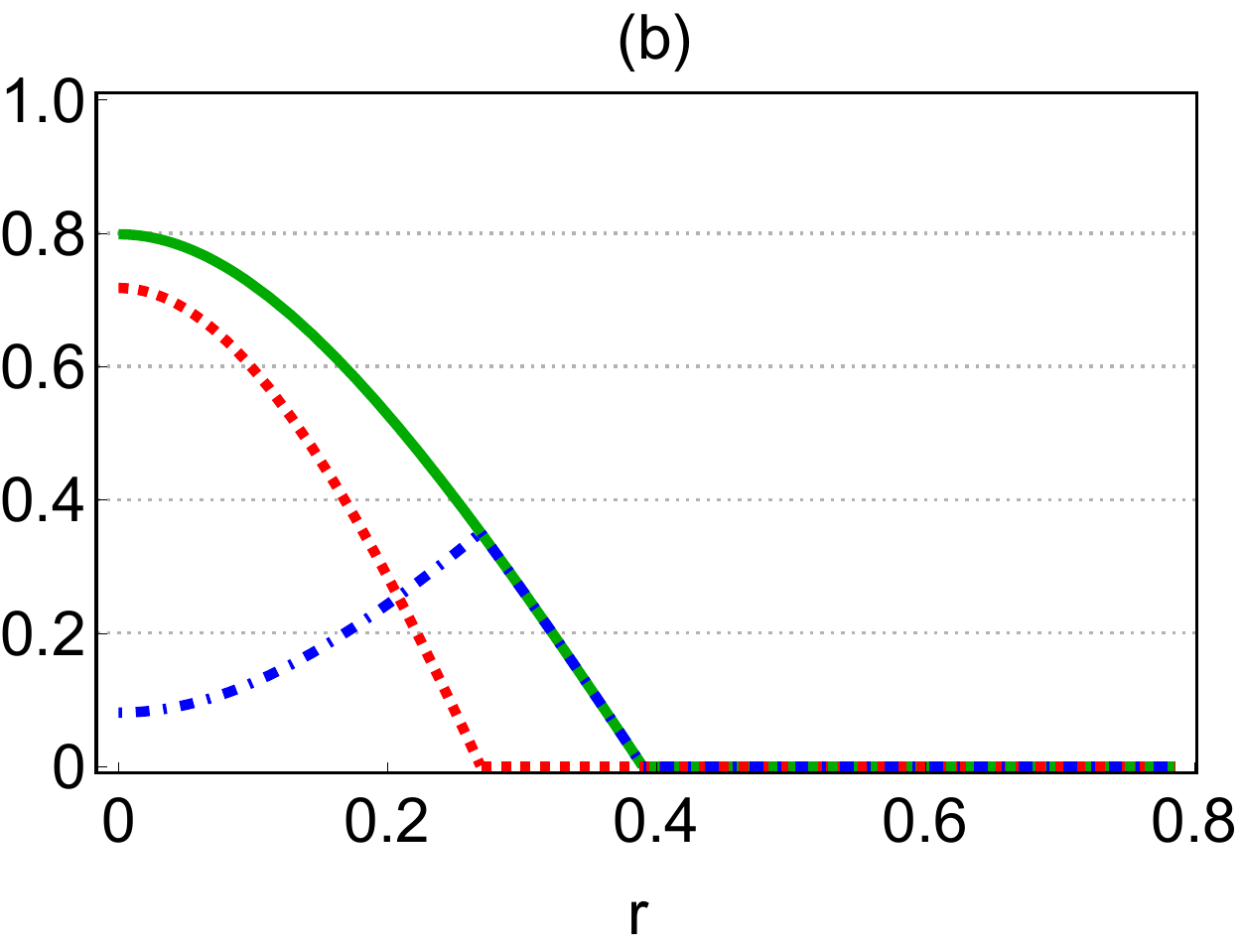}
	\includegraphics[width=0.315\linewidth, height=3.5cm]{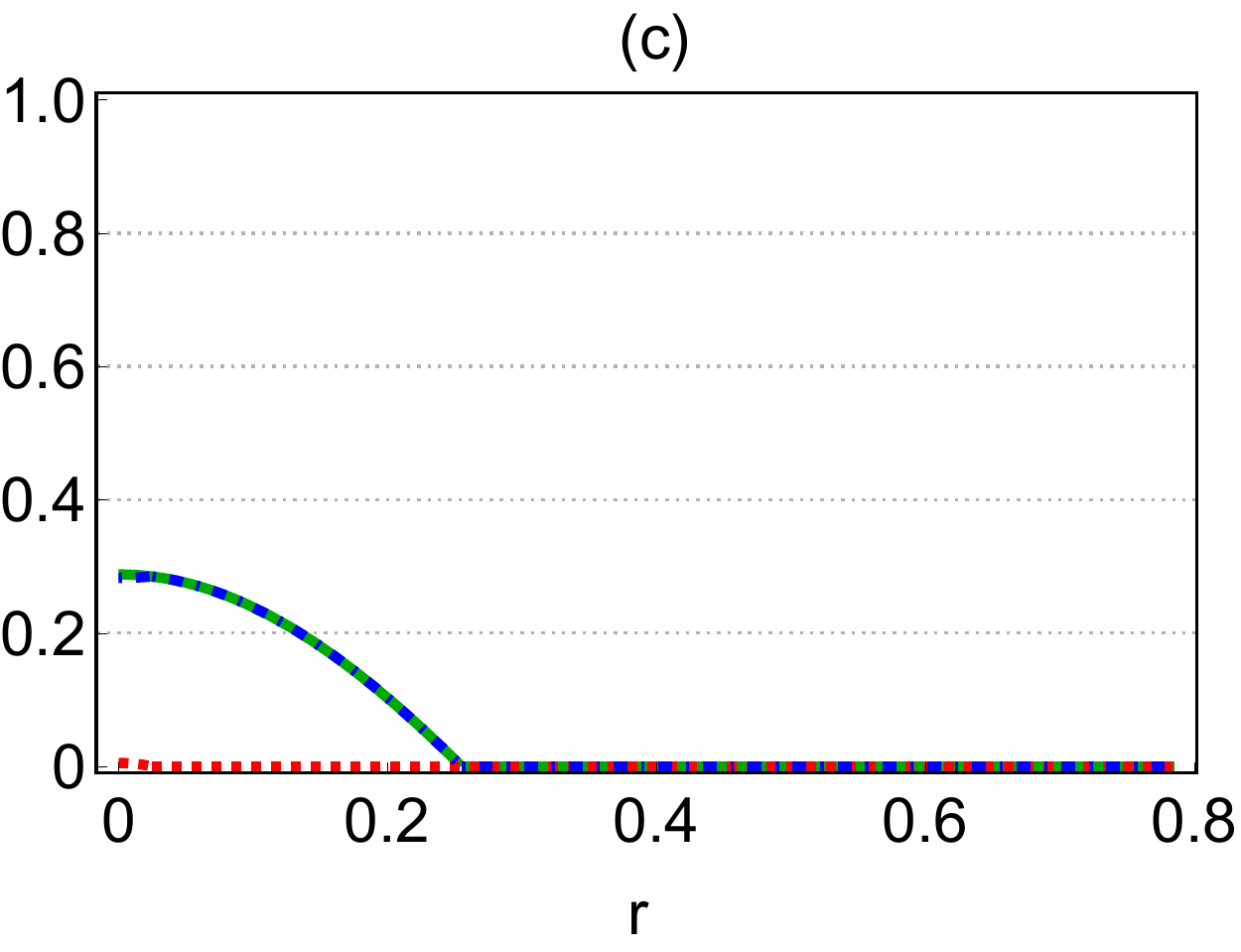}
	\caption{The same as Fig.(\ref{figDS-q}), but only  the qutrit is accelerated.}
	\label{figDS-qt}
\end{figure}

 Fig.(\ref{figDS-qt}) displays the behavior of the steerability degrees $\mathcal{S}_{AB}$ and $\mathcal{S}_{BA}$, when only the qutrit is accelerated.  The behavior is similar to that displayed in Fig.(\ref{figDS-q}), where the degree of steerabilities decrease as the decoherence parameters increase. However, the possibility that the qubit steers the qutrit is larger than that shown when  the qutrit steers the qubit, namely $\mathcal{S}_{AB}>\mathcal{S}_{BA}$. Moreover, the sudden decreasing phenomenon of the steerabilities is depicted, while the gradual decay is displayed in Fig.(\ref{figDS-q}). The steerability from the qutrit  to the qubit vanishes at smaller values of the acceleration parameter.  The difference between the degree of steerability $|\mathcal{S}_{AB}-\mathcal{S}_{BA}|$ is larger than that displayed in Fig.(\ref{figDS-q}), where only the qutrit is accelerated.

 From Figs.(\ref{figDS-q}) and (\ref{figDS-qt}), it is clear that, if the qubit or the qutrit are accelerated the $\mathcal{S}_{AB}>\mathcal{S}_{BA}$. The possibility that the  qubit  steers the qutrit  survives at  large  values of the acceleration, while that  depicted for the qutrit vanishes at small accelerations. The degree of steerability  decreases as the mixing parameter increases.

\section{Conclusion}\label{Con}
In this article, we investigate the steering process for an accelerated  qubit -qutrit  system.  The initial system depends on a mixing parameter, which is considered as a decoherence parameter. Due to the acceleration process, there is an additional decoherence  depends on which subsystem is accelerated.  Therefore, we quantify the decoherence rate that rises from the initial state settings and from the acceleration process. Moreover, the non-locality, as well as, the amount of quantum correlations are quantified by using the local quantum uncertainty. Finally, we discuss the possibility that the qubit/qutrit steers each other besides we  quantify the degree of steerability.

Our results  show  that, at the absences of the mixing parameter, namely the initial state is maximally entangled state, the purity  of the qubit-qutrit system is  maximum. However, as one increases the mixing parameter, the decoherence of the total qubit-qutrit system  and the qutrit subsystems  increase gradually. On the other hand,  the decoherence  could be improved if the qubit is accelerated with large accelerations.  Moreover, the decoherence arises from the acceleration process, where it increases as the acceleration increases. The decoherence rate  depends on the accelerated subsystem, where the decay rate that is displayed when the qutrit is accelerated is larger than that displayed if  the qubit is accelerated.

Due to the decoherence, the accelerated qubit-qutrit system loses its  coherence and consequently the amount of the  non-classical correlation  decreases. The local quantum uncertainty is used to quantify the  quantum correlations, where at small values of the mixing parameter, it  decreases as the acceleration increases. However, the amount of quantum correlations increases at large values of the mixing and acceleration parameters.  This behavior of these correlations is exhibited  clearly when both subsystems are accelerated.

The possibility that each subsystem steers the other is investigated at different initial state settings.  It is clear that, the steerability decreases as the mixing and acceleration parameters increase. The predicted steerability that the small size subsystem steers the large size subsystem is larger than that displayed for the large subsystem steers the small subsystem. The degree of steerability decreases gradually if  only the qubit is accelerated, while it vanishes  suddenly,  when the qutrit or both subsystems are accelerated.  Moreover, the qubit has the ability to steers the qutrit at large accelerations. Furthermore, when the qutrit is accelerated the degree of steerability vanishes at small accelerations. The difference between the degrees of steerability  increases when the qutrit or both subsystems are accelerated.

Finally, for this family of the initial state, the acceleration process   improves the purity of  the initial state and its subsystems.
 The steerability between different sizes of accelerated subsystems is possible. The degree of steerability depends on the initial state settings and the values of the acceleration parameter.

\end{document}